\documentclass[aps,prd,twocolumn,superscriptaddress,showpacs]{revtex4}
\usepackage{graphicx}

\begin{document}

\title{High Statistics Search for the ${\it \mathnormal{\Theta}}^+$(1.54) Pentaquark State}

\affiliation{Institute of Physics, Academia Sinica, Taipei 11529, Taiwan, Republic of China}
\affiliation{University of California, Berkeley, California 94720}
\affiliation{Fermi National Accelerator Laboratory, Batavia, Illinois 60510}
\affiliation{Universidad de Guanajuato, 37000 Le\'{o}n, Mexico}
\affiliation{Illinois Institute of Technology, Chicago, Illinois 60616}
\affiliation{Universit\'{e} de Lausanne, CH-1015 Lausanne, Switzerland}
\affiliation{Lawrence Berkeley National Laboratory, Berkeley, California 94720}
\affiliation{University of Michigan, Ann Arbor, Michigan 48109}
\affiliation{University of South Alabama, Mobile, Alabama 36688}
\affiliation{University of Virginia, Charlottesville, Virginia 22904}

\author{M.J. Longo\footnote{Corresponding author, email:~mlongo@umich.edu}}
\affiliation{University of Michigan, Ann Arbor, Michigan 48109}

\author{R.A. Burnstein}
\affiliation{Illinois Institute of Technology, Chicago, Illinois 60616}

\author{A. Chakravorty}
\affiliation{Illinois Institute of Technology, Chicago, Illinois 60616}

%\author{A. Chan}
%\affiliation{Institute of Physics, Academia Sinica, Taipei 11529, Taiwan, Republic of China}

\author{Y.C. Chen}
\affiliation{Institute of Physics, Academia Sinica, Taipei 11529, Taiwan, Republic of China}

\author{W.S. Choong}
\affiliation{University of California, Berkeley, California 94720}
\affiliation{Lawrence Berkeley National Laboratory, Berkeley, California 94720}

\author{K. Clark}
\affiliation{University of South Alabama, Mobile, Alabama 36688}

\author{E.C. Dukes}
\affiliation{University of Virginia, Charlottesville, Virginia 22904}

\author{C. Durandet}
\affiliation{University of Virginia, Charlottesville, Virginia 22904}

\author{J. Felix}
\affiliation{Universidad de Guanajuato, 37000 Le\'{o}n, Mexico}

\author{Y. Fu}
\affiliation{Lawrence Berkeley National Laboratory, Berkeley, California 94720}

\author{G. Gidal}
\affiliation{Lawrence Berkeley National Laboratory, Berkeley, California 94720}

%\author{P. Gu}
%\affiliation{Lawrence Berkeley National Laboratory, Berkeley, California 94720}

\author{H.R. Gustafson}
\affiliation{University of Michigan, Ann Arbor, Michigan 48109}

%\author{C. Ho}
%\affiliation{Institute of Physics, Academia Sinica, Taipei 11529, Taiwan, Republic of China}

\author{T. Holmstrom}
\affiliation{University of Virginia, Charlottesville, Virginia 22904}

\author{M. Huang}
\affiliation{University of Virginia, Charlottesville, Virginia 22904}

\author{C. James}
\affiliation{Fermi National Accelerator Laboratory, Batavia, Illinois 60510}

\author{C.M. Jenkins}
\affiliation{University of South Alabama, Mobile, Alabama 36688}

\author{T. Jones}
\affiliation{Lawrence Berkeley National Laboratory, Berkeley, California 94720}

\author{D.M. Kaplan}
\affiliation{Illinois Institute of Technology, Chicago, Illinois 60616}

\author{L.M. Lederman}
\affiliation{Illinois Institute of Technology, Chicago, Illinois 60616}

\author{N. Leros}
\affiliation{Universit\'{e} de Lausanne, CH-1015 Lausanne, Switzerland}

\author{F. Lopez}
\affiliation{University of Michigan, Ann Arbor, Michigan 48109}

\author{L.C. Lu}
\affiliation{University of Virginia, Charlottesville, Virginia 22904}

\author{W. Luebke}
\affiliation{Illinois Institute of Technology, Chicago, Illinois 60616}

\author{K.B. Luk}
\affiliation{University of California, Berkeley, California 94720}
\affiliation{Lawrence Berkeley National Laboratory, Berkeley, California 94720}

\author{K.S. Nelson}
\affiliation{University of Virginia, Charlottesville, Virginia 22904}

\author{H.K. Park}
\affiliation{University of Michigan, Ann Arbor, Michigan 48109}

\author{J.-P. Perroud}
\affiliation{Universit\'{e} de Lausanne, CH-1015 Lausanne, Switzerland}

\author{D. Rajaram}
\affiliation{Illinois Institute of Technology, Chicago, Illinois 60616}

\author{H.A. Rubin}
\affiliation{Illinois Institute of Technology, Chicago, Illinois 60616}

%\author{P.K. Teng}
%\affiliation{Institute of Physics, Academia Sinica, Taipei 11529, Taiwan, Republic of China}

\author{J. Volk}
\affiliation{Fermi National Accelerator Laboratory, Batavia, Illinois 60510}

\author{C.G. White}
\affiliation{Illinois Institute of Technology, Chicago, Illinois 60616}

\author{S. White}
\affiliation{Illinois Institute of Technology, Chicago, Illinois 60616}

\author{P. Zyla}
\affiliation{Lawrence Berkeley National Laboratory, Berkeley, California 94720}

\collaboration{HyperCP Collaboration}
\noaffiliation

\date{\today}

\begin{abstract}
We have searched for $\mathnormal{\Theta}^+(1.54) \rightarrow K^0 p$ decays using data from the 1999 run of the HyperCP experiment at Fermilab.  We see no evidence for a narrow peak in the $K_S^0p$ mass distribution near 1.54~GeV/$c$ among 106\,000 $K_S^0p$ candidates, and obtain an upper limit for the fraction of $\mathnormal{\Theta}^+(1.54)$ to $K_S^0p$ candidates of $<$0.3\% at 90\% confidence. 
\end{abstract}


\maketitle

\section{Introduction}

A number of experiments have recently reported a narrow peak in the $K^+n$ and $K^0_Sp$ mass spectra near 1.54 GeV/$c^2$ \cite{Nakano,barth,kubarovsky,barmin,asratyan,stepanyan,aleev, airapetian,aslanyan,chekanov,abdel}.  Negative searches have also been reported \cite{Bai,knopfle,alt,wang,christian, blume,wengler,pinkenburg,lafferty,coleman,salur}.  A resonance in the $K^+n$ system with strangeness +1 and baryon number +1 would require five quarks as constituents.  The $K^0_Sp$ system, on the other hand,  is ambiguous since the strangeness could be either $+1$ requiring the system to be ``exotic",  or $-1$ like the normal hyperons.  The presumed pentaquark state is referred to as the $\mathnormal{\Theta}^+$.  The width of the peak seems to be narrower than the experimental mass resolutions, which vary between about 9 and 27 MeV/$c^2$.  However, Cahn and Trilling \cite{Cahn} argue that the intrinsic width would have to be $<$1 MeV/$c^2$ based on the data of Barmin {\em et al.} \cite{barmin} and $K^+d$ cross section data.  This is far narrower than any previously discovered hadronic resonance, and therefore the dynamics of the $\mathnormal{\Theta}^+$ decay would have to be ``exotic", as well as its quantum numbers.  On the other hand, the production of $\mathnormal{\Theta}^+$, which involves the rearrangement of quarks, just as in any other hadronic production process, should not be exotic.  Thus, for example, a virtual {\em N*} produced diffractively from an incident proton could decay into a $\mathnormal{\Theta}^+$ and $\overline{K}$, and a $K^+$ scattering off a neutron could produce $\mathnormal{\Theta}^+ X^0$.  The general expectation is that the inclusive $\mathnormal{\Theta}^+$ production cross sections should be fairly large, especially for $K^+$ beams \cite{Liu}, and approximately energy independent well above threshold, just as for any other hadronic state.

\section{The HyperCP Experiment}

\begin{figure}[b]
\includegraphics[width=3.375in]{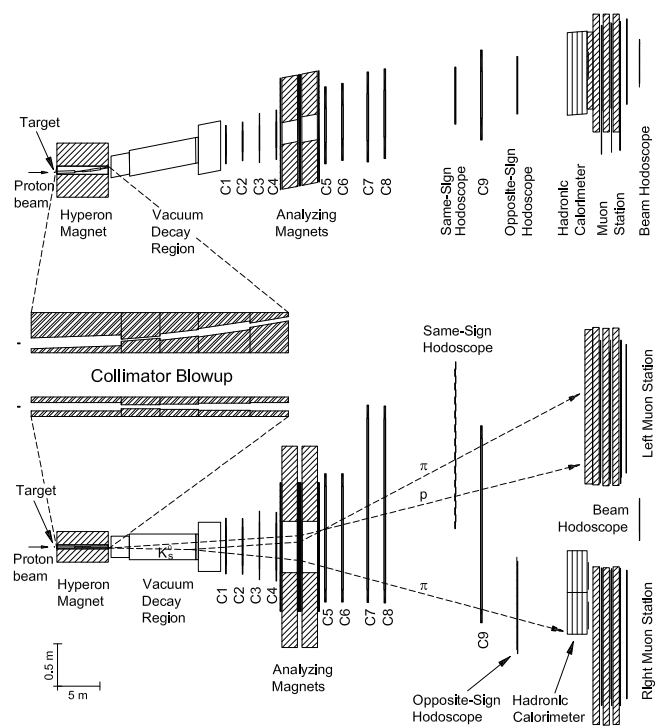}
\caption{Elevation (top) and plan (bottom) views of the HyperCP spectrometer with blowups of the collimator.}
\label{fig:spect}
\end{figure}

The HyperCP detector (Fig.~\ref{fig:spect}) is described in detail elsewhere \cite{burnstein}.  It was designed principally to investigate {\em CP} violation in $\Xi^-/\overline{\Xi}^+$ decays.  A charged secondary beam was produced at 0$^\circ$ by 800~GeV protons interacting in a copper target.  A collimator channel was embedded in a 6.0-meter-long dipole magnet with a field of 1.667~T that deflected the secondary beam upward at an angle of 19.5~mrad.  The exit aperture of the collimator was 2~cm wide by 1~cm high made with machined tungsten blocks.  The collimator selected a broad momentum band ranging from about 120 to 220~GeV/$c$.  When a positively charged secondary beam was selected, the beam was a mixture of mainly protons and $\pi^+$, with about 5\% $K^+$, and a smaller fraction of positively charged hyperons.  Most of the long-lived particles produced in the target passed through an evacuated pipe (Vacuum Decay Region) and encountered only the thin windows of the Vacuum Decay Region and wire chambers that were in the beam, while most hyperons decayed in the Vacuum Decay Region or upstream of it.  However, over half of the triggers recorded were from events produced in the tungsten near the exit of the defining collimator.  Events produced there were used in the pentaquark search.

The HyperCP spectrometer employed two dipole magnets (Analyzing Magnets) with a total bending power of 4.72 T m.  The Analyzing Magnets were preceded by four wire chambers and followed by five wire chambers.  Each chamber had two planes of vertical wires and planes with wires inclined at angles of $+26.6^\circ$ and $-26.6^\circ$ to the vertical.  The wire pitches ranged from 1~mm for the upstream chambers to 2~mm for the larger downstream chambers.  The momentum resolution of the spectrometer was 0.44\% at 100~GeV/$c$ \cite{burnstein}.  The spectrometer and data acquisition \cite{daq} were designed for high rates, and in the course of the experiment, $2{\times}10^{11}$ triggers were recorded on magnetic tape.  The majority of the running was done using a positive-polarity secondary beam in order to obtain a sufficiently large sample of $\overline{\Xi}$$^+$ decays for the {\em CP} violation test.  The trigger used for the pentaquark search required a coincidence between hits in scintillation counter hodoscopes on either side of the beam.   This trigger was prescaled by a factor of 100, so that only 1\% of the potential candidates were recorded.

The experiment did not have particle identification.  However, $K^0_S\rightarrow \pi^+\pi^-$ decays could be cleanly identified by reconstructing the two-pion mass and requiring that it be consistent with the $K^0$ mass.   Charged tracks with $>$ 50\% of the total momentum had a very high probability of being protons.  Thus the pentaquark search was restricted to the decay $\mathnormal{\Theta}^+ \rightarrow K^0_S p$.

\section{Data Analysis}

For the pentaquark search, events were selected with vertices originating from the lower lip of the tungsten collimator.  Cuts on the vertex position along the beam direction were applied to select events originating in the last 20~cm of the collimator. The vertical position of the vertex was required to be within 4.0~mm of the top edge of the lower lip of the collimator.  Events were rejected if their total momentum vector traced back to the production target.  Events with three or more tracks were chosen with the total momentum of all tracks between 120 and 250 GeV/$c$.  Events containing $\Lambda \rightarrow p\pi^-$ or $K^+\rightarrow \pi^+ \pi^+ \pi^-$ decays were rejected.  Cuts were also made to reject ``ghost" tracks which are near-duplicate tracks generated by the track-fitting program. (Events with ghost tracks from $\Lambda\rightarrow p\pi^-$ decays that also fit the $K_S^0p$ hypothesis were found to give a narrow peak at 1.54 GeV/$c^2$ in the $K^0_Sp$ mass \cite{ghost}).

\begin{figure}[b]
%figure2
\begin{center}
\includegraphics[width=3.0in]{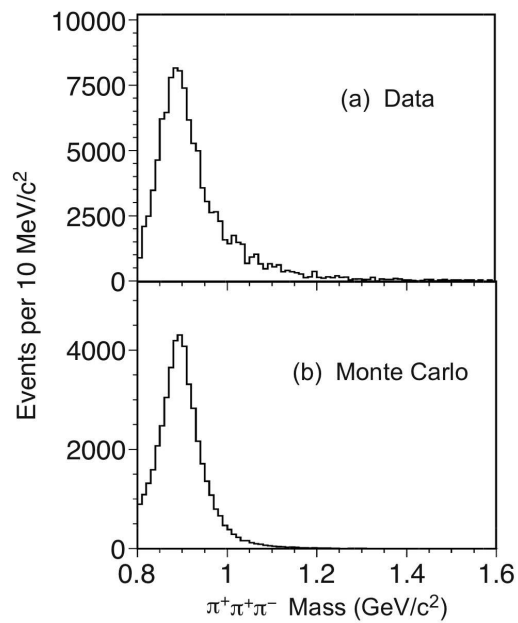}
\end{center}
\caption{(a) $K^*$(0.892) signal obtained by subtracting the $K^0_{\rm sideband} \pi$ data from the $K^0\pi$ data.  (b)  Monte Carlo simulation of the $K^*$(0.892) produced in the last 20~cm of the collimator.  The simulation assumes a Breit-Wigner shape with a full width $\Gamma$ = 51~MeV.}
\label{fig:k_star}
\end{figure}

Note that the high beam momentum and broad momentum spread in the beam are advantages in the pentaquark search because the mechanism proposed by Dzierba {\em et al.} \cite{dzierba} by which narrow enhancements in the $K\,N$ mass distribution can be produced through kinematic reflections of mesons, such as $f_2(1.275), a_2(1.320)$ and $\rho_3(1.690)$, would not be effective.  Also, the kinematic reflection mechanism for producing false mass peaks involving $K^0$ and $\Lambda$ decay overlap, as suggested by Zavertyaev \cite{zavert}, would not work. 

Events with $K_S^0$'s were selected by requiring that the two lower momentum particles have opposite sign and reconstructing their mass under the $\pi^+ \pi^-$ assumption.    Events with $K^0_S$'s were used to search for $K^*(0.892) \rightarrow K^0_S\pi$ decays that were produced in the collimator.  Figure~\ref{fig:k_star} compares the $K^*$(0.892) signal with a Monte Carlo simulation.  The $K^*$ signal was isolated by subtracting the $\pi^+\pi^+\pi^-$ mass distribution for events with a $\pi^+ \pi^-$ mass in the $K^0_S$ mass sidebands from the $K^0_S \pi$ mass distribution.  A similar subtraction was used to isolate the  $\Sigma$(1.385) resonance peak in $\Lambda \pi$ events;  this also agreed with Monte Carlo expectations.

\begin{figure}[t]
%figure3
%\includegraphics[width=3.0in]{penta_Figure3.eps}
\includegraphics[width=3.0in]{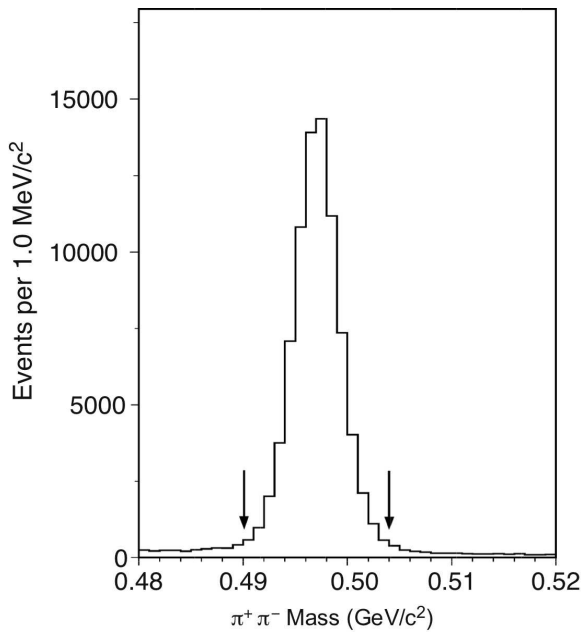}
\caption{Reconstructed $\pi^+ \pi^-$ mass for the events in the $\mathnormal{\Theta}^+$ search sample.  The cuts to select $K^0_S\rightarrow \pi^+\pi^-$ are shown.}
\label{fig:ks}
\end{figure}

\begin{figure}[b]
%figure4
%\includegraphics[width=3.0in]{penta_Figure4.eps}
\includegraphics[width=3.0in]{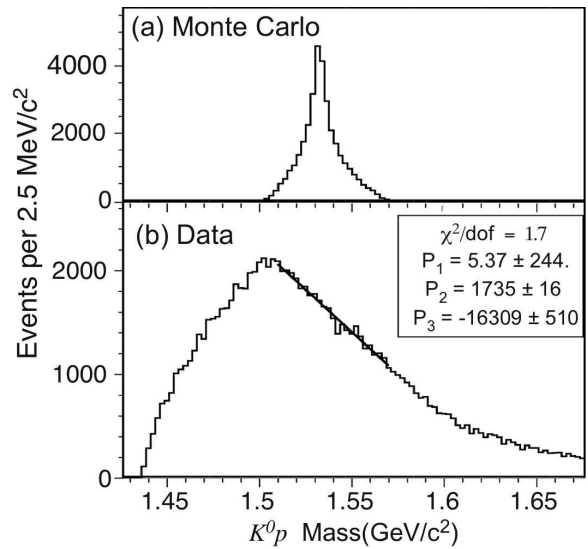}
\caption{The $K_S^0p$ invariant mass distribution.  (a) Monte Carlo events with the $\mathnormal{\Theta}$(1.54) produced in the last 20~cm of the collimator.  (b) Reconstructed  $K_S^0p$ mass for the total pentaquark candidate sample.  The smooth curve between 1.51 and 1.57 GeV/$c^2$ is a fit to the shape in the top plot centered at 1.53 GeV/$c$ with normalized amplitude $P_1$, plus a linear background, $P_2 + P_3\times$ (Mass).}
\label{fig:ks_p}
\end{figure}

The  reconstructed $\pi^+ \pi^-$ mass for the events in the pentaquark sample is shown in Fig.~\ref{fig:ks}.  The reconstructed $K^0_Sp$ mass is shown in Fig.~\ref{fig:ks_p}, where we have combined the positive and negative beam data samples, and both $\mathnormal{\Theta}^+\rightarrow K^0_S p$ and $\overline{\mathnormal{\Theta}}$$^- \rightarrow K^0_S \overline{p}$ candidates are included.  With a cut requiring the proton momentum fraction to be $>$0.50, we estimate the non-proton contamination in the $K^0_S p$ sample to be $\sim$1\%.  The upper plot in Fig.~\ref{fig:ks_p} shows the mass distribution expected from a Monte Carlo simulation of a narrow $\mathnormal{\Theta}^+(1.54)\rightarrow K^0_S p$.  The expected rms mass resolution is about 11~MeV/$c^2$ with a non-gaussian distribution.  In the total sample of 106\,000 $K^0_Sp$ and $K^0_S\bar{p}$ candidates there is no sign of a peak near 1.54~GeV/$c^2$.

The smooth curve overlaying the histogram in the 1.51--1.57 GeV/$c^2$ range is a fit to the data with the shape in the upper plot centered at 1.53~GeV/$c^2$ plus a linear background.  This shows the maximum number of events in the putative $\mathnormal{\Theta}^+(1.53)$ peak to be 317 at 90\% confidence level.  Thus we find that at most 0.3\% of the $K^0_Sp$ candidates could come from $\mathnormal{\Theta}^+$ decays.  By comparison, Barmin {\em et al.} \cite{barmin} reported 29 events in their $\mathnormal{\Theta}^+$ peak among 541 $K^0_Sp$ events with a similar mass spectrum in the DIANA bubble chamber experiment.  Because of the large momentum spread and mixed beam composition, as well as the unknown production spectrum, it was not possible to estimate limits for ${\mathnormal{\Theta}}^+$ production cross sections.

\section{Discussion}

The main differences between our experiment and those that reported $\mathnormal{\Theta}^+(1.54)$ peaks are the higher beam energy, a mixed beam containing $p, \pi^+, K^+,$ and charged hyperons with a broad momentum spread, as well as better mass resolution and far more events.  There is no fundamental reason why the ratio of $\mathnormal{\Theta}^+$ to overall $K^0p$ production should decrease significantly at higher beam energies.  To produce a pentaquark, a beam proton has to pick up a $d \bar{s}$ quark pair from the sea.  To produce a $\overline{\Xi}^+,~\bar{u}, \bar{s}, \bar{s}$ quarks need to be created.  Naively one would expect $\mathnormal{\Theta}^+$ production to be much larger than $\overline{\Xi}^+$.  Yet the data sample used in the pentaquark search contained over 75\,000 $\overline{\Xi}^+$ decays and 70\,000 $\Xi^-$ decays that originated from the lower lip of the collimator.  Thus the absence of substantial $\mathnormal{\Theta}^+$ production in our data is extremely puzzling.

\begin{acknowledgments}
We wish to thank Alex Dzierba for useful conversations.  The authors are indebted to the staffs of Fermilab and the participating institutions for their vital contributions.  This work was supported by the U.S. Department of Energy and the National Science Council of Taiwan, R.O.C.~~E.C.D. and K.S.N. were partially supported by the Institute for Nuclear and Particle Physics.  K.B.L. was partially supported by the Miller Institute for Basic Research Science.
\end{acknowledgments}


\begin{thebibliography}{99}
\bibitem{Nakano} T.~Nakano {\em et al.}, Phys. Rev. Lett. {\bf 91}, 012002 (2003).

\bibitem{barth} J.~Barth {\em et al.}, Phys. Lett. B {\bf 572}, 127 (2003).

\bibitem{kubarovsky} V.~Kubarovsky {\em et al.}, Phys. Rev. Lett. {\bf 92}, 032001 (2004).

\bibitem{barmin} V.~V.~Barmin {\em et al.}, Phys. Atom. Nucl. {\bf 66}, 1715 (2003).

\bibitem{asratyan} A.~Asratyan, A.G.~Dolgolenko, and M.A.~Kubantsev, Phys. Atom. Nucl. {\bf 67,} 682 (2004); hep-ex/0309042.

\bibitem{stepanyan} S.~Stepanyan {\em et al.}, Phys. Rev. Lett. {\bf 91}, 252001 (2003).

\bibitem{aleev} A.~Aleev {\em et al.}, hep-ex/0401024.

\bibitem{airapetian} A.~Airapetian {\em et al.}, Phys. Lett. B {\bf 585}, 213 (2004).

\bibitem{aslanyan} P.~Aslanyan {\em et al.}, hep-ex/0403044.

\bibitem{chekanov} S.~Chekanov {\em et al.}, Phys. Lett. B {\bf 591}, 7 (2004).

\bibitem{abdel} M.~Abdel-Bary {\em et al.}, Phys. Lett. B {\bf 595}, 127 (2004).

\bibitem{Bai} J.~Z.~Bai {\em et al.}, Phys. Rev. D {\bf 70}, 012004 (2004).

\bibitem{knopfle} K.~T.~Knopfle {\em et al.}, J. Phys. {\bf G30}:S1363 (2004).

\bibitem{alt} I.~Alt {\em et al.}, Phys. Rev. Lett. {\bf 92}, 042003 (2004).

\bibitem{wang} M.-J.~Wang, Quarks \& Nuclear Physics Conference, Bloomington, 2004; http://www.qnp2004.org/.

\bibitem{christian} D.~Christian, Quarks \& Nuclear Physics Conference, Bloomington, 2004; http://www.qnp2004.org/.

\bibitem{blume} C.~Blume, www.ikf.physik.uni-frankfurt.de/na49/collab-2001/Collaborations\_Meeting\_12\_2003/Blume/Blume\_coll-dec-03-penta.pdf.

\bibitem{wengler} T.~Wengler, hep-ex/0405080.

\bibitem{pinkenburg} C.~Pinkenburg, J. Phys. G30:S1201 (2004).

%\bibitem{lafferty} G.~Lafferty, www.hep.man.ac.uk/u/gdl/xmas.ppt;

%\bibitem{lafferty} P.~Hansen, XII Intl. Workshop on Deep Inelastic Scattering, Pleso, April 2004; http://www.saske.sk/dis04/.
\bibitem{lafferty} S. Schael {\em et al.}, Phys. Lett. B {\bf 599}, 1 (2004).

%\bibitem{coleman} J.~Coleman, presented at the 2004 APS April meeting, http://www.slac.stanford.edu/BFROOT/.
\bibitem{coleman} B.~Aubert {\em et al.}, hep-ex/0408064.

\bibitem{salur} S.~Salur, nucl-ex/0403009.

\bibitem{Cahn} R.~Cahn and G.~Trilling, Phys. Rev. D {\bf 69}, 011501(2004).

\bibitem{Liu} W.~Liu and C.~M.~Ko, Phys. Rev. C {\bf 68}, 045203 (2003).

\bibitem{burnstein} R.A.~Burnstein {\em et al.,} hep-ex/0405034 (to be published in Nucl. Instrum. Methods).

\bibitem{daq} C.G. White {\em et al.}, Nucl.\ Instrum.\ Meth. A {\bf 474}, 67 (2001).

\bibitem{ghost} M. Longo, Quarks \& Nuclear Physics Conference, Bloomington, 2004; http://www.qnp2004.org/.

\bibitem{dzierba} A.~R.~Dzierba {\em et al.}, Phys. Rev. D {\bf 69}, 051901 (2004). 

\bibitem{zavert} M.~Zavertyaev, hep-ph/0311250.
\end{thebibliography}
\end{document}